\documentclass[3p,times,procedia]{elsarticle}
\usepackage{nupha_ecrc}

\volume{00}

\firstpage{1}

\journalname{Nuclear Physics A}

\runauth{}

\jid{nupha}

\jnltitlelogo{Nuclear Physics A}

\usepackage{amssymb}

\usepackage[figuresright]{rotating}

\begin{document}

\begin{frontmatter}



\dochead{}

\title{Chiral Magnetic Effect in Heavy Ion Collisions}


\author[ad1,ad2]{Jinfeng Liao}

\address[ad1]{ Physics Department and Center for Exploration of Energy and Matter,
Indiana University, 2401 N Milo B. Sampson Lane, Bloomington, IN 47408, USA.} 
\address[ad2]{RIKEN BNL Research Center, Bldg. 510A, Brookhaven National Laboratory, Upton, NY 11973, USA.}

\begin{abstract}
The Chiral Magnetic Effect (CME) is a remarkable phenomenon that stems from highly nontrivial interplay of QCD chiral symmetry, axial anomaly, and gluonic topology.  It is  of fundamental importance to search for the CME in experiments. The heavy ion collisions provide a unique environment where a hot chiral-symmetric quark-gluon plasma  is created, gluonic topological fluctuations generate chirality imbalance, and very strong magnetic fields $|\vec{\bf B}|\sim m_\pi^2$ are present during the early stage of such collisions. Significant efforts have been made to look for CME signals in heavy ion collision experiments. In this contribution we give a brief overview on the status of such efforts.
\end{abstract}

\begin{keyword}
Chiral Magnetic Effect \sep Heavy Ion Collision \sep Chiral Anomaly \sep QCD Topology 

\end{keyword}

\end{frontmatter}


\section{Introduction}
\label{sec:1}

Our modern society and our daily lives heavily rely upon {\it electricity}, the knowledge about which originated from physicists' curiosity in exploring and understanding {\it electromagnetism of matter}.  External electric and magnetic fields were used early on, and are still widely used today in many ways, as probes for the properties of materials under investigation. One of the most important discoveries, is the Ohm's Law, $\vec {\bf J} = \sigma \vec{\bf E}$, i.e. the generation of an electric current $\vec {\bf J}$ in the presence of external electric field $\vec{\bf E}$ with the conductivity coefficient $\sigma$ characterizing the charge transport property of matter. Coming to the discussion of a new state of matter, the quark-gluon plasma (QGP), one might  imagine the remote possibility of {\it quarkicity} and wonder what would be the electromagnetic transport properties of QGP when probed by external  electromagnetic (EM) fields. The quarks are carriers of electric charges that respond to external EM fields. However the quarks are carriers of color charges as well and their transport dynamics would be dominated by strong interaction in QGP. The question of EM transport properties in QGP therefore concerns interplay between QED and QCD, and becomes more interesting particularly due to the {\it chiral symmetry} of QCD.

 With the light quark current masses negligible compared to relevant energy scales,  the QCD Lagrangian has an approximate chiral symmetry. However this symmetry is not realized in the ground state due to  the formation of quark-anti-quark pair condensate in the vacuum of QCD which is dominated by strong quantum fluctuations with nonperturbative interactions. This is of course the well-known phenomenon of {\it spontaneously broken chiral symmetry} which is a fundamental property of QCD.  On the other hand, first principle computations from lattice QCD have shown that with increasing temperature such vacuum condensate will be ``melted'' away with the chiral symmetry being restored at sufficiently high temperature $T>T_c\sim 165\rm MeV$~\cite{Borsanyi:2010bp,Bazavov:2011nk}. Therefore, {\it a chiral-symmetric QGP at high temperature is also a fundamental prediction of QCD}. With such symmetry of light quark sector, the currents in QGP can be independently examined for right-handed (RH) and left-handed (LH) quarks. The chiral currents $\vec{\bf J}_{R/L}$ can then be combined into the familiar vector current $\vec{\bf J}= \vec{\bf J}_{R} + \vec{\bf J}_{L}$ and axial current  $\vec{\bf J}_5= \vec{\bf J}_{R} - \vec{\bf J}_{L}$. Recent studies on the  transport properties of QGP in responses to external EM fields have found the generalized ``Ohm's Table'' for QGP: 
\begin{eqnarray}
\left(\begin{array}{c} \vec{\mathbf J} \\ \vec{\mathbf
    J}_5 \end{array}\right) = \left(\begin{array}{cc} \sigma &
  \sigma_5   \\ \sigma_{\chi e}   & \sigma_s
\end{array}\right) \left(\begin{array}{c}
\vec{\mathbf E} \\ \vec{\mathbf B}
\end{array}\right). \label{eq_vaeb}
\end{eqnarray}
In addition to the familiar Ohmic conducting effect, three interesting new effects are identified in the above table:  the Chiral Magnetic Effect (CME)~\cite{Kharzeev:2004ey,Kharzeev:2007tn,Kharzeev:2007jp}, the Chiral Separation Effect (CSE)~\cite{son:2004tq,Metlitski:2005pr}, and the Chiral Electric Separation Effect (CESE)~\cite{Huang:2013iia,Jiang:2014ura}. For detailed discussions on these anomalous transport effects and for more complete reference lists, we refer the readers to recent reviews in e.g.~\cite{Kharzeev:2015znc,Kharzeev:2013ffa,Liao:2014ava}. In this contribution, we focus on the Chiral Magnetic Effect and related phenomena in the quark-gluon plasma. 

The CME predicts the generation of a vector current $\vec{\bf J}$ in response to external magnetic field $\vec{\bf B}$: 
\begin{eqnarray}\label{cme}
\vec{\bf J} =\sigma_5 \vec{\bf B}
\end{eqnarray} 
One may immediately realize that {\it normally} the above transport process is forbidden by symmetry considerations: $\vec{\bf J}$ is $\cal P$-odd and $\cal CP$-even, while $\vec{\bf B}$ is $\cal P$-even and $\cal CP$-odd. Indeed the CME could occur only in {\it a $\cal P$- and $\cal CP$-odd environment} where the CME conductivity $\sigma_5$ (being a pseudo-scalar quantity) is nonzero. More specifically $\sigma_5=C_A \mu_5$ with $C_A$ a constant. The $\mu_5$ is a chiral chemical potential that quantifies the presence of {\it chirality imbalance} i.e. the difference between the numbers of RH and LH chiral fermions in the system. For a chiral QGP with nonzero $\mu_5\neq0$, a CME current is generated along the direction of   $\vec{\bf  B}$. Intuitively the CME may be understood in the following way. The magnetic field leads to a spin
polarization (i.e. ``magnetization'') effect, with quarks' spins
preferably aligned along the $\vec{\mathbf B}$ field direction, which
implies $\langle \vec s \rangle \propto (Qe) \vec{\mathbf B}$. Quarks
with specific chirality have their momentum $\vec p$ direction
correlated with spin $\vec s$ orientation: $\vec p \, || \vec s$ for RH
quarks, while $\vec p \, || (-\vec s)$ for LH ones. In the presence of
chirality imbalance, i.e. $\mu_5\neq 0$, there will be a net
correlation between average spin and momentum $\langle \vec p \rangle
\propto \mu_5 \langle \vec s \rangle$.   It is therefore evident that $\langle \vec p \rangle \propto (Qe)
\mu_5 \vec{\mathbf B}$, which implies a vector current of these quarks
$\vec{\mathbf J} \propto \langle \vec p \rangle \propto (Qe) \mu_5
\vec{\mathbf B}$.

There remains however the crucial question of how to achieve a chiral QGP with chirality imbalance in the first place.  The process of creating nonzero chirality pertains to the famous {\it ``chiral anomaly''} (or often called ``triangle/axial anomaly''). This anomaly implies the breaking of axial current conservation by quantum loop effect, $\partial_\mu J^\mu_5 = C_A \vec{\bf E}\cdot \vec{\bf B}$. Therefore chirality imbalance  can be created through   nonzero $\vec{\bf E}\cdot \vec{\bf B}$: such fields can be either QED EM fields or QCD chromo-EM fields. 

In QCD the gluon configurations with globally nonzero $\vec{\bf E}\cdot \vec{\bf B}$ are known to exist: they are the {\it topological solitons such as instantons and sphalerons}. These topological objects play crucial roles in our understanding of nonperturbative dynamics in QCD vacuum~\cite{'tHooft:1999au} as well as in hot QGP~\cite{Shuryak:2014zxa,Sharma,Xu:2014tda}. Such configurations have their global topological winding number  $Q_w\sim \int \vec{\bf E}\cdot \vec{\bf B}$, and when coupled with chiral fermions, can generate definite amount of global chirality imbalance via chiral anomaly: $N_R - N_L =2 Q_w$ for each flavor of light quarks.  In addition there are also strong local fluctuations  of topological charge density (i.e. locally nonzero chromo $\vec{\bf E}\cdot \vec{\bf B}$) from the initial conditions in the glasma (see e.g.~\cite{Hirono:2014oda,Fukushima:2015tza}) which would similarly translate into local fluctuations of chirality imbalance.  In short, the chirality imbalance is a direct manifestation of the QCD topological fluctuations and could become observable through the CME current.

Summarizing the introductory discussions, the CME is a remarkable phenomenon that stems from highly nontrivial interplay of QCD chiral symmetry, axial anomaly, and gluonic topology. The pertinent CME conductivity $\sigma_5=C_A \mu_5$ has distinctive features by virtue of its topological origin: it is $\cal T$-even and thus non dissipative, its constant $C_A$ is completely fixed from anomaly relations and thus takes universal value from weak to strong coupling and encompassing various physical systems.  It is therefore of fundamental importance to search for the CME in experiments. The heavy ion collisions provide a unique environment where a hot chiral-symmetric QGP is created, gluonic topological fluctuations generate chirality imbalance, and very strong magnetic fields $|\vec{\bf B}|\sim m_\pi^2$ are present during the early stage of such collisions. Significant efforts have been made to look for CME signals in heavy ion collision experiments. In the rest of this contribution we give a brief overview on the status of such efforts, with an emphasis on most recent developments.

\section{In Search of Chiral Magnetic Effect}

In heavy ion collisions the fast-moving ions with sizable positive charges create extremely strong magnetic  fields $\vec{\bf B}$ at early time, with the azimuthal orientation pointing in the out-of-plane direction (see e.g. detailed discussions in \cite{Huang:2015oca,Bloczynski:2012en,HYL}). Given a nonzero initial axial charge $\mu_5$,
the CME current $\vec{\bf J}$ is induced along  $\vec{\bf B}$ (thus in the
out-of-plane direction) with its sign depending upon $\mu_5$. Specifically an
electric current $\vec{\bf J}_Q$ forms as a result of all
contributing quark-level CME currents. This current transports positive charges toward one pole of the QGP
fireball and negative charges toward the opposite pole, thus forming a
dipole moment in the charge distribution of the QGP.  Such a charge separation in the out-of-plane direction can be parameterized in the paricle
  azimuthal distribution in a form~\cite{Voloshin:2004vk}: 
\begin{eqnarray}
\frac{dN_{\pm}}{d\phi} \propto 1 + 2v_1\cos(\phi-\Psi_{\rm RP}) + 
2v_2\cos[2(\phi-\Psi_{\rm RP})] + ... + 2a_{\pm}\sin(\phi-\Psi_{\rm RP}) + ...,
\end{eqnarray} 
where  
$v_1$ and $v_2$ are coefficients accounting for the so-called directed
and elliptic flow. The CME-induced charge separation signal lies in the $\sin(\phi-\Psi_{\rm RP})$ term with the coefficient  $a_+ = - a_- \propto  \mu_5 |\vec{\bf B}|$. 
The $\Psi_{\rm RP}$ denotes reaction plane angle and for simplicity we will set   $\Psi_{\rm RP}=0$ for the rest of our discussions. 

The complication here is that the chirality imbalance $\mu_5$ arises from fluctuations and changes its sign from event to event. Upon average over many events the $<\sin\phi>\sim <\mu_5 |\vec{\bf B}|>$ term (which itself is a $\cal P$-odd term) would vanish. This is not surprising as there is no global parity violation in QCD. The next best thing one could try, is to measure the {\it event-by-event fluctuations} $\sim <( \mu_5 |\vec{\bf B}|)^2>$ of such a $\cal P$-odd term. This could be achieved by measuring the   azimuthal correlations for same-charge and opposite-charge  pairs~\cite{Voloshin:2004vk}: 
\begin{eqnarray}
 \label{gamma_ob}
\gamma_{\alpha\beta}= < \cos(\phi_i+\phi_j)>_{\alpha\beta}\, , \qquad
\delta_{\alpha\beta}= < \cos(\phi_i-\phi_j)>_{\alpha\beta}\, ,
\end{eqnarray}
with $\alpha,\beta=\pm 1$ labeling the hadrons'  charge and $\phi_{i,j}$  the azimuthal angles of two final state charged hadrons.  One could easily construct the reaction-plane-projected correlations from the above, $<\cos\phi_i \cos\phi_j>_{\alpha\beta}=(\gamma_{\alpha\beta}+\delta_{\alpha\beta})/2 $ and $<\sin\phi_i \sin\phi_j>_{\alpha\beta}=(\gamma_{\alpha\beta}-\delta_{\alpha\beta})/2 $ with the latter containing the specific CME contribution we are looking for. However, these observables (themselves being $\cal P$-even) suffer from possible non-CME background contributions, i.e. $<\cos\phi_i \cos\phi_j>_{\alpha\beta}= <v_1^2> + B_{in}$ and $<\sin\phi_i \sin\phi_j>_{\alpha\beta}= <a_\alpha a_\beta> + B_{out}$. The rationale to measure the $\gamma$-correlator  is to reduce the background contributions~\cite{Voloshin:2004vk}, i.e.  $\gamma_{\alpha\beta}= <v_1^2>  - <a_\alpha a_\beta>+ (B_{in} - B_{out})$ with  the term $<v_1^2>$   known to be negligible and with the hope that $(B_{in} - B_{out})$ would also be negligible due to cancellation. A pure CME signal would give $<\cos\phi_i \cos\phi_j>^{CME}_{\alpha\beta}=0$ and $<\sin\phi_i \sin\phi_j>^{CME}_{\alpha\beta}\propto (\alpha \beta) <( \mu_5 |\vec{\bf B}|)^2>$, implying $\gamma^{CME}_{\alpha\beta} = - \delta^{CME}_{\alpha\beta} \propto (-\alpha \beta) <( \mu_5 |\vec{\bf B}|)^2>$.  

These observables were   measured by STAR collaboration~\cite{STAR_LPV} and later by PHENIX~\cite{PHENIX_LPV} at RHIC as well as by ALICE~\cite{ALICE_LPV} at LHC.  While the initial data appeared to be in line with qualitative features of CME, it was quickly realized that there are still sizable background contributions that are completely ``orthogonal'' to CME expectation~\cite{Bzdak:2009fc}. The issue is that the  $(B_{in} - B_{out})$ term in $\gamma$-correlator is {\it not negligible}: its contribution, at the level of elliptic flow $v_2$ (that quantifies the in-plane/out-of-plane difference in bulk evolution), is comparable to measured correlations. Several possible sources such as transverse momentum conservation (TMC) and local charge conservation (LLC) effects were identified: for detailed discussions and further references on elliptic-flow-driven background issues see e.g. \cite{Kharzeev:2015znc,Flow_CME}.    

The above complicated situation during the first stage (roughly 2004$\sim$2010) of search for CME in heavy ion collisions, cried out for much more hard work in order to reach a conclusion. On the experimental side, new measurements and analyses would be needed to help separate flow-driven backgrounds from desired CME signal and to find clues of other possible CME-related phenomena. On the theoretical side, what's badly needed would be the realistic and quantitative modelings of CME physics based on state-of-the-art hydrodynamic simulations that incorporate anomalous transport and account for background contributions.

\section{Recent Developments}

The past several years have seen a new stage of CME search with exciting progress on a number of front, including the theoretical foundations, the separation of backgrounds, the quantitative modelings, new ideas of anomalous transport phenomena, etc. In this Section we briefly review these new developments. 

Before discussing the CME search in heavy ion collisions, let us first mention that a whole new territory for the experimental realization of CME has recently been opened in condensed matter physics. In systems called Dirac and Weyl semimetals, effective chiral fermions are found to exhibit the Chiral Magnetic Effect: we refer the readers to Q.~Li's contribution in these proceedings for details~\cite{QL}.

\subsection{Building Theoretical Foundations}

The bulk evolution of the QGP fireball created in heavy ion collisions is well described by hydrodynamics. In order to develop quantitative simulations for CME, one needs to understand how the macroscopic hydrodynamics for matter with chiral fermions may be modified to account for microscopic quantum anomaly. A new theoretical framework, the {\it anomalous hydrodynamics}, has been developed recently~\cite{Son:2009tf}. For a chiral current $J^\mu$ the hydrodynamic equation gets modified as $\partial_\mu J^\mu= C_A E_\mu B^\mu$ due to anomaly. Recall that in usual hydrodynamics, to the leading order of gradient expansion beyond ideal hydrodynamics, the constituent relation for the current would be fixed entirely by the second law of thermodynamics $\partial_\mu s^\mu\ge 0$ i.e. the entropy could only grow in time. The same line of argument, in the presence of anomaly, leads to a remarkable result~\cite{Son:2009tf}: the leading viscous term $\nu_\mu$ is required to include two new contributions that are precisely the chiral magnetic current and the chiral vortical current. The development of anomalous hydrodynamics  paves the way for quantitative simulations of anomalous transport in heavy ion collisions. Early attempts for applying such a framework were made in \cite{Hongo:2013cqa,Yee:2013cya}.  
 
 Given that anomaly effect does modify hydrodynamics (which is an effective description of many-body system near equilibrium), it is tempting to ask how  the microscopic anomaly manifests itself in the out-of-equilibrium situation e.g. in the kinetic framework.  Like hydrodynamics, the kinetic theory is another ``effective description'' at a different level of ``coarse-graining''. A number of important theoretical results have been obtained recently, developing toward a matured framework called {\it ``chiral kinetic theory''}~\cite{Stephanov:2012ki,Son:2012wh,Chen:2012ca,Chen:2014cla,Chen:2015gta,MS}. While the usual kinetic theory is based on classical description of quasiparticles, the chiral kinetic theory accounts for $\hat{O}(\hbar)$ quantum effect from the spin-momentum correlation for a chiral fermion.  This is achieved by modifying the classical equation of motion with an anomalous velocity term from the so-called Berry curvature in momentum space. Such a framework correctly reproduces the chiral anomaly relation as well as various known anomalous transport effects. This new versatile theory provides the foundation for future modeling of out-of-equilibrium CME contribution at early time of a heavy ion collision when the magnetic field is the strongest.

\subsection{Charge Separation Measurements: Fighting the Backgrounds}

As previously mentioned, the most pressing issue in the current charge separation measurements is how to separate the elliptic flow driven backgrounds from the magnetic field driven effect. A number of possibilities were proposed. For example one may examine the same correlation measurements for   different colliding systems, such as UU~\cite{STAR_LPV_UU} or CuAu~\cite{HYL,Deng:2014uja,Niida} collisions, that have different patterns of geometric eccentricity and electromagnetic fields in contrast with AuAu collisions.

\begin{figure}[hbt!]
\includegraphics[width=0.32\textwidth,height=0.25\textwidth]{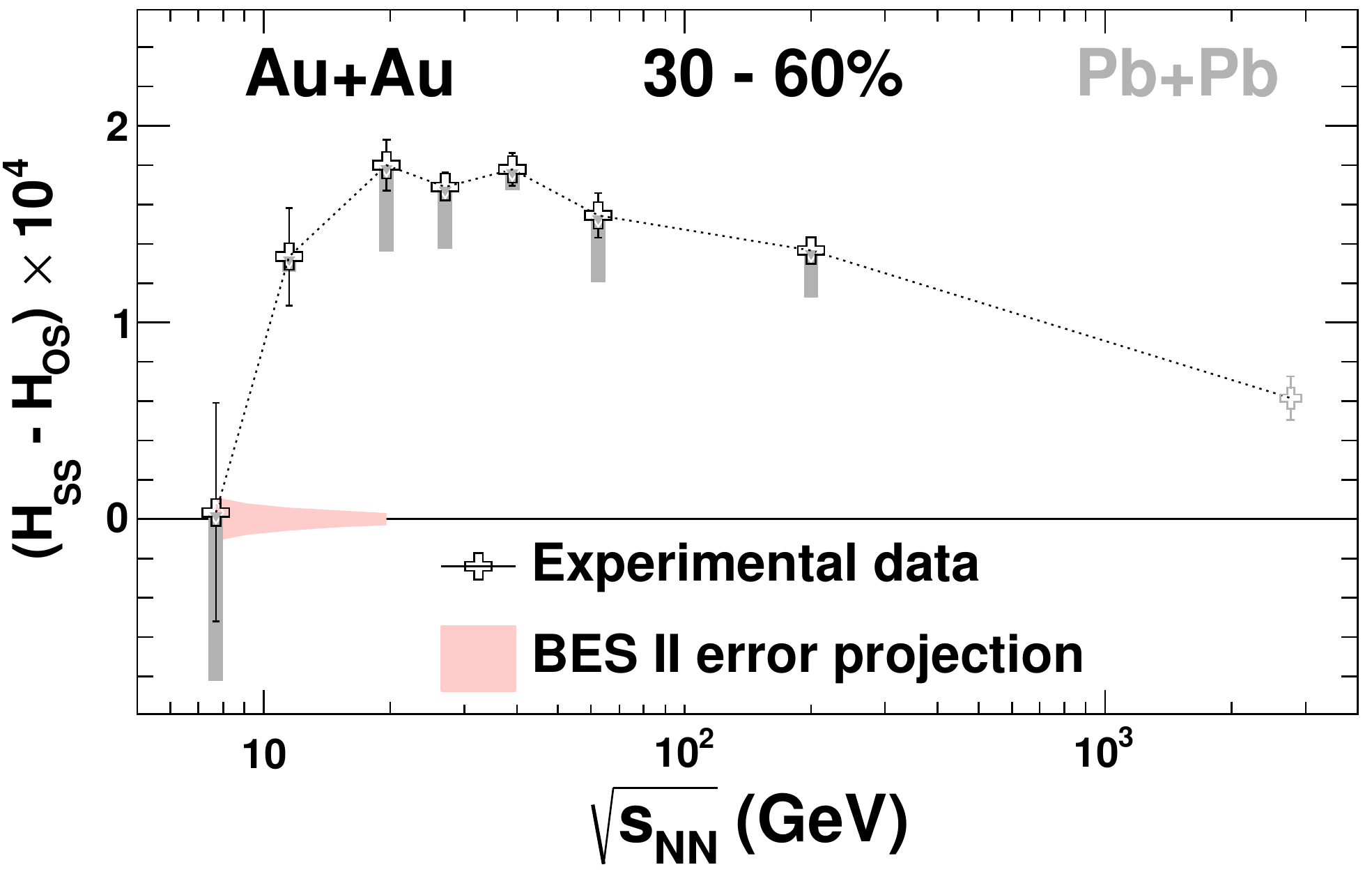}
\includegraphics[width=0.32\textwidth,height=0.32\textwidth]{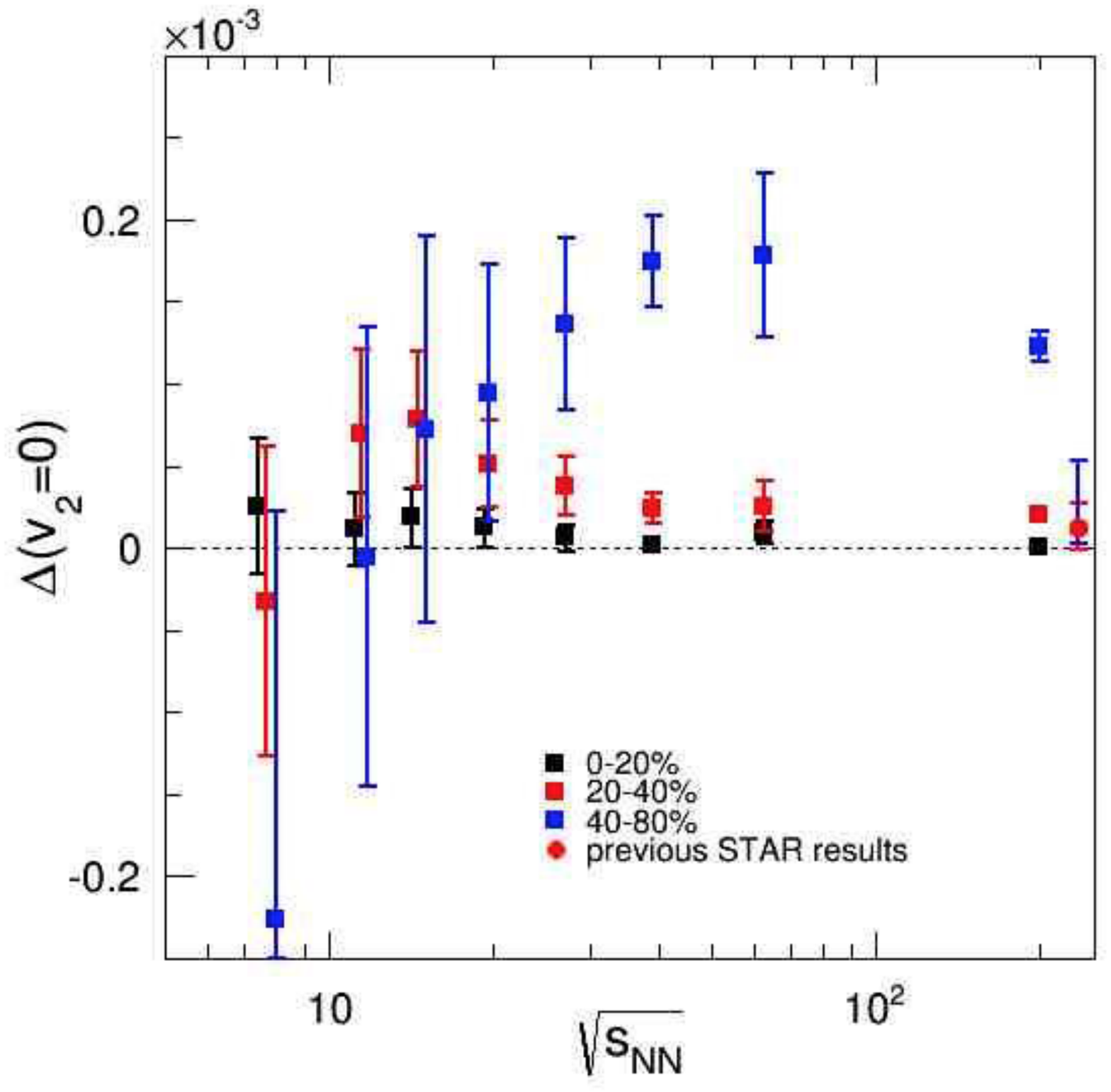}
\includegraphics[width=0.32\textwidth,height=0.34\textwidth]{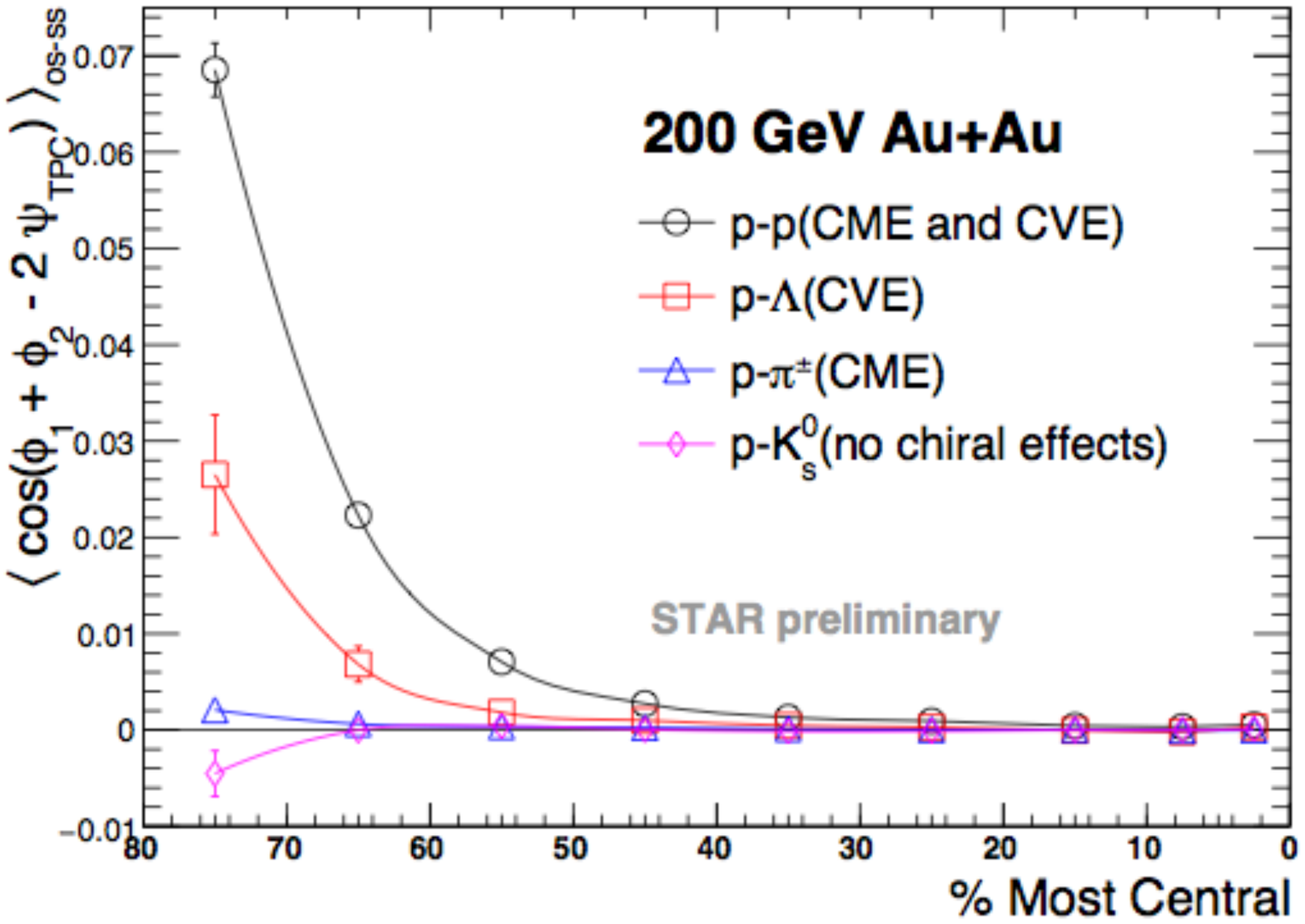}
\caption{(Color online)
 STAR data for extracted charge-dependent correlations after separating flow-driven backgrounds with two different methods, from~\cite{STAR_LPV_BES} (left) and from~\cite{STAR_LPV4,STAR_Tu} (middle). 
(Right) $\gamma$-correlators for identified hadron pairs, from~\cite{STAR_Wen}. 
\vspace{-0.05in}
\label{fig_data}
}
\end{figure}

A different strategy, suggested in \cite{Flow_CME}, is to examine and separate the component in $\gamma$ and $\delta$ correlations that may bear direct dependence on elliptic flow $v_2$. Motivated by study of known background effects like TMC and LLC, one may postulate the following two-component scenario:  
  \begin{eqnarray}
 \label{gamma_ob_2}
\gamma_{\alpha\beta}= \kappa \, v_2 \, F_{\alpha\beta} - H_{\alpha\beta} \, , \qquad
\delta_{\alpha\beta}= F_{\alpha\beta} + H_{\alpha\beta} \, ,
\end{eqnarray}
where $F$ represents the flow-driven background strength while $H$ represents the potential CME signal. The parameter $\kappa$ is a constant of order one, to account for acceptance corrections. Using data for $\gamma$, $\delta$ and $v_2$ one can thus extract the $F$ and $H$ signals. Recently STAR has performed such analysis with Beam Energy Scan (BES-I) data~\cite{STAR_LPV_BES}: see Fig.~\ref{fig_data} (left). The so-extracted signal $H$ shows a clear charge dependence  with nontrivial trends in beam energy, and provides a very encouraging evidence in support of CME interpretation.   

Another interesting analysis was also performed by STAR collaboration, with  similar spirit as the above but a different method~\cite{STAR_LPV4,STAR_Tu}.  The idea is to {\it bin the events in  a given centrality class according to their elliptic flow $v_2$} and see how the charge separation signal (which is similar but slightly different from the $\gamma$-correlator) shifts as a function of the $v_2$ bin values. A linear dependence on $v_2$ is found, and the so-extracted intercept $\Delta(v_2=0)$ is arguably the desired signal that is free of flow backgrounds: see results in Fig.~\ref{fig_data} (middle). The advantage of this method is that no assumption of a free parameter like $\kappa$ in (\ref{gamma_ob_2}) is needed. As one can see, the signal from this analysis appears to be {nonzero and consistent with the previous analysis.} Possible further background sources and event-plane resolution corrections are under investigation. 
\vspace{-0.5in}

Taken together, these recent analyses aiming at separating flow backgrounds and extracting possible CME signal, have demonstrated a strong case for establishing and quantifying the CME in heavy ion collisions.  Special opportunity for deciphering CME signal versus flow-driven backgrounds may be further provided by performing {\it isobaric collisions experiments} at RHIC. One particular proposal   is to compare $^{96}\, _{44}Ru + ^{96}\, _{44}Ru$ and $^{96}\, _{40}Zr + ^{96}\, _{40}Zr$ collisions: the two colliding systems should create nearly identical fireball evolution (including anisotropic flows) albeit with $\sim 10\%$ variation in $\vec{\bf B}$ field strength (i.e.  $\sim 20\%$ variation in  CME signal). Current measurement capability shall either prove or falsify such variation.

\subsection{Toward Quantitative Modelings}

  To draw a definitive conclusion, it is vital to develop anomalous hydrodynamic 
 simulations that quantify the CME signals with realistic initial
 conditions as well as account for background
 contributions. Two recent works~\cite{Hirono:2014oda,Yin:2015fca} have made significant first steps toward this goal. 
 \vspace{-0.1in}
 
\begin{figure}[hbt!]
\begin{center}
\includegraphics[width=0.5\textwidth,height=0.4\textwidth]{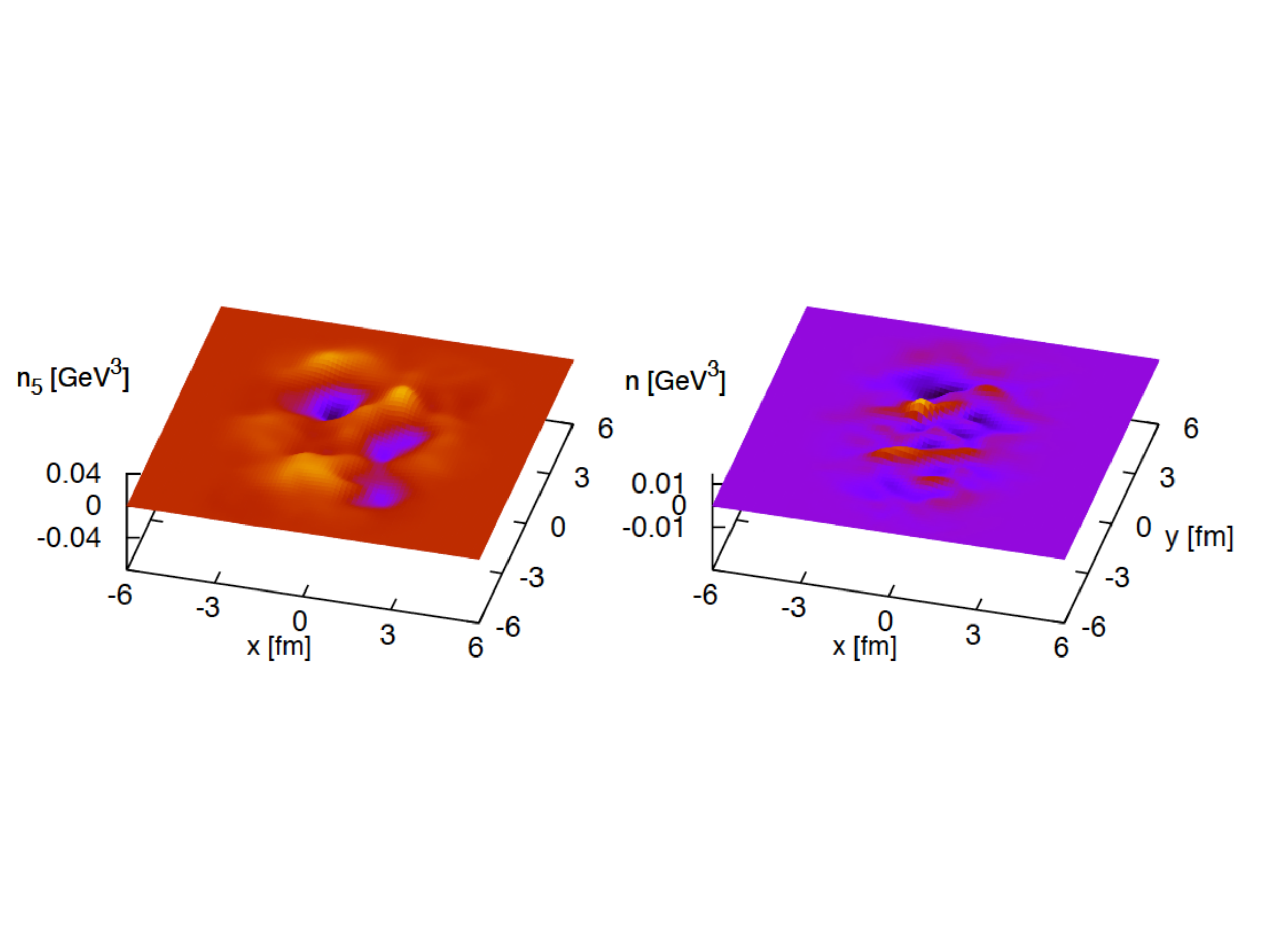}
\includegraphics[width=0.45\textwidth]{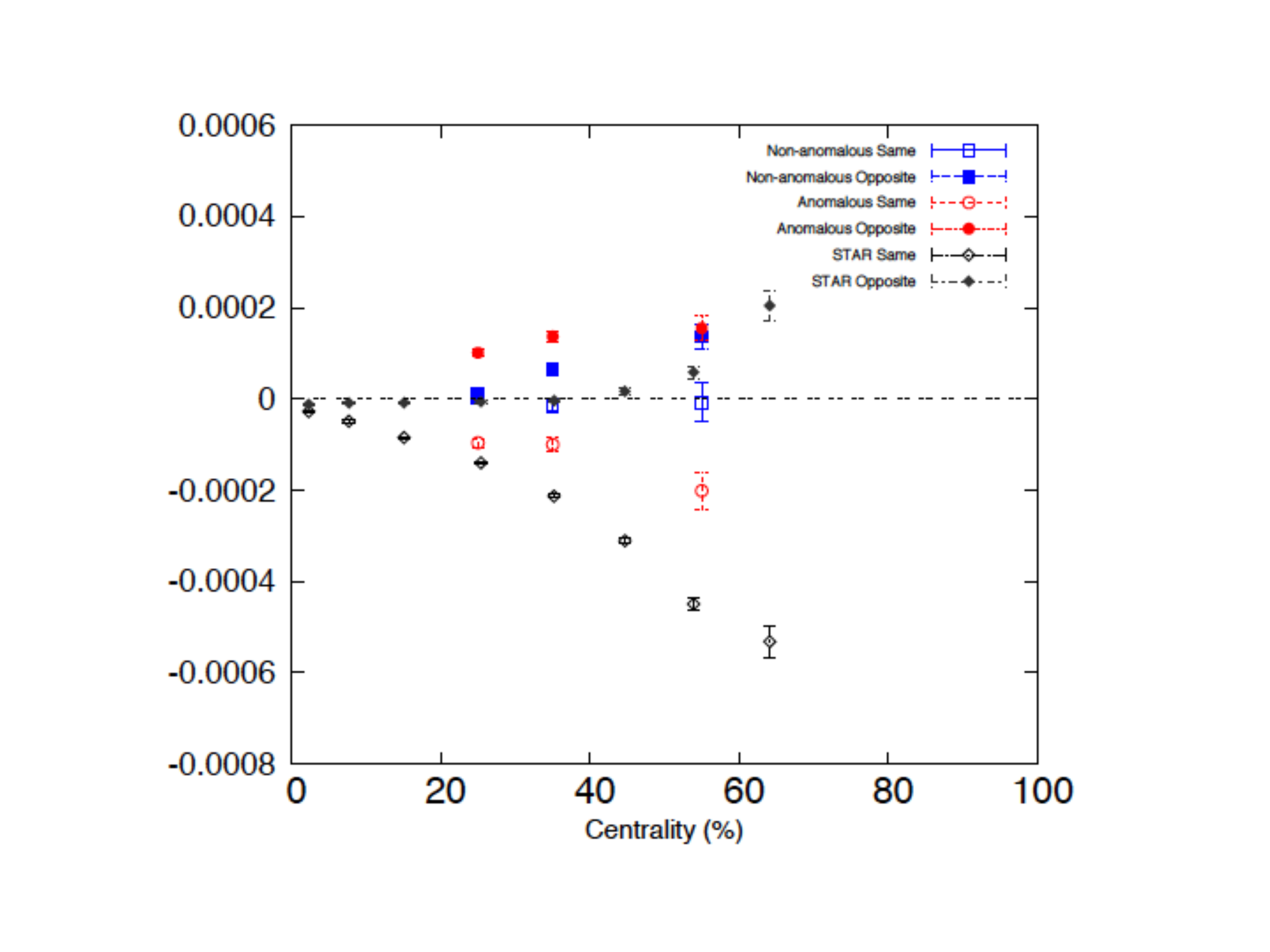}
\end{center} \vspace{-0.45in}
\caption{(Color online)
Distributions of the chiral (left) and electric (middle) charge densities in the transverse plane at mid-rapidity and a proper time $\tau = 1.5$fm/c of a Au+Au collision event at $\sqrt{s_{\rm NN}}=200$ GeV from event-by-event  anomalous hydrodynamics simulations. (Right) The $\gamma$-correlators for same and opposite charge pairs computed from the same simulations.  (See \cite{Hirono:2014oda} for details.) \label{fig_HHK}. 
}\vspace{-0.1in}
\end{figure}

  In \cite{Hirono:2014oda}, the authors have performed the
 first event-by-event simulations of the CME in the anomalous
 hydrodynamic framework with   initial conditions describing topological fluctuations in  glasma flux
 tubes at the early stage of a heavy collision. In Fig.~\ref{fig_HHK} the snapshots of the resulting  chiral (left) and electric (middle) charge densities for one AuAu collision event are shown. The computed final hadron charge asymmetries with anomalous transport are compared with the same observables computed without anomalous terms, as shown in Fig.~\ref{fig_HHK} (right). The results suggest that the final charge asymmetries are sensitive to the presence of anomaly effect and the obtained CME signals agree with data within a factor of two (which leaves room for possible background contributions not included in this calculation).

\begin{figure}[hbt!]
\includegraphics[width=0.32\textwidth,height=0.25\textwidth]{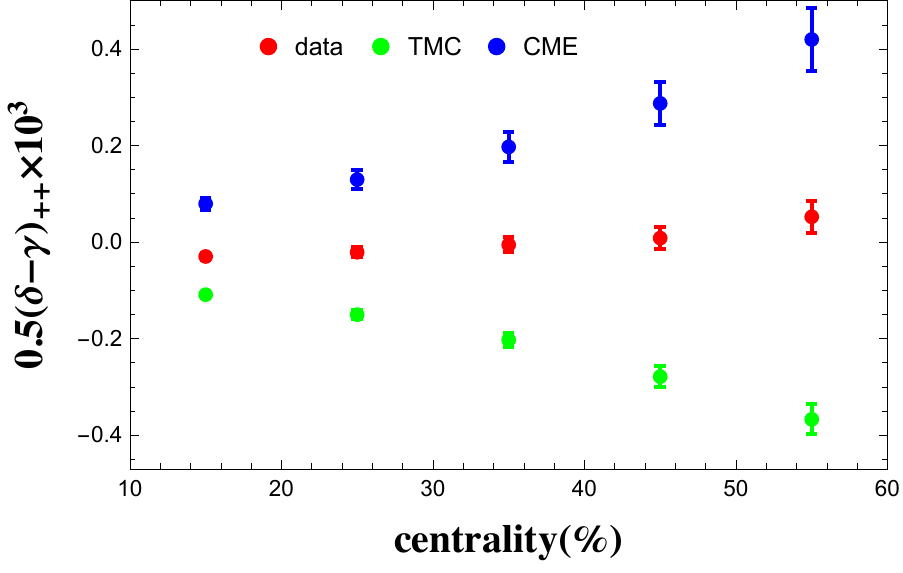}
\includegraphics[width=0.32\textwidth,height=0.25\textwidth]{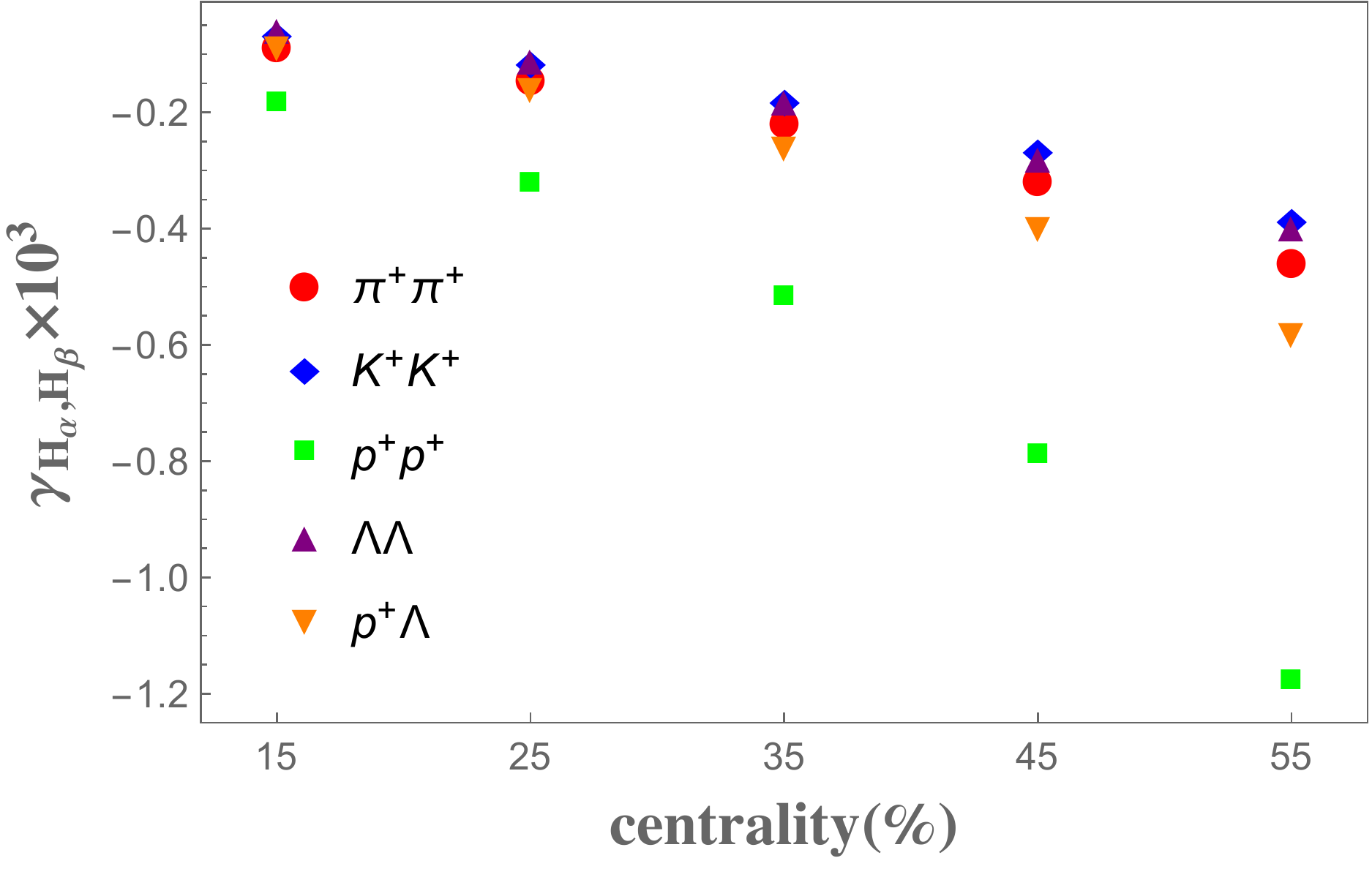}
\includegraphics[width=0.32\textwidth,height=0.25\textwidth]{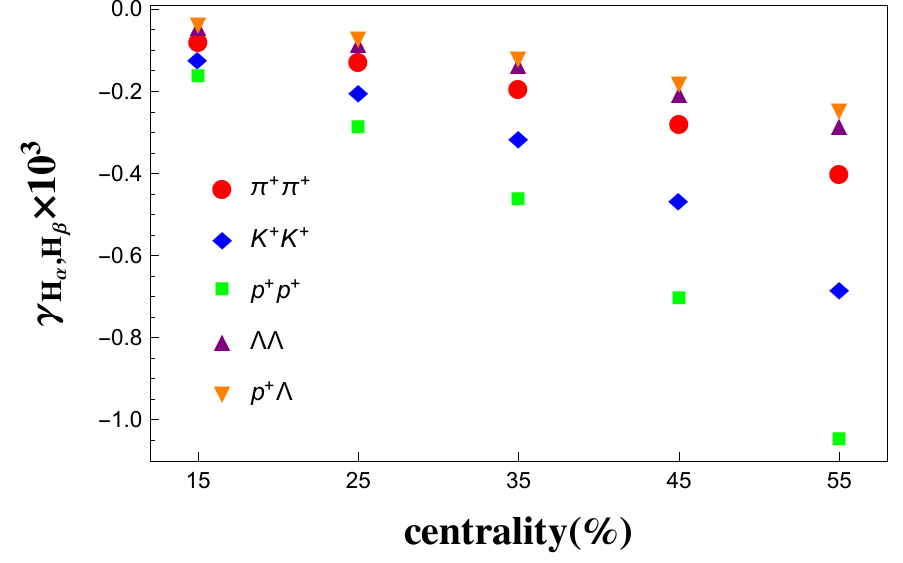}
\caption{(Color online)
 Same-charge correlation data for $(\delta-\gamma)/2$  and its decomposition into CME and TMC contributions for different centrality (left). 
Predictions for $\gamma$-correlators of various identified same-charge hadrons versus centrality for two-flavor case (middle) and three-flavor (right) case. (See \cite{Yin:2015fca} for details.)\vspace{-0.1in}
\label{fig_YL}
}
\end{figure}

 In \cite{Yin:2015fca}, the authors have made the first attempt
 to consistently quantify contributions to observed charge
 correlations from both the CME signal and background contributions in one and same
 framework that integrates anomalous hydro with data-validated bulk
 viscous hydro simulations. The anomalous hydrodynamic equations for pertinent currents are solved in a linearized way on top of the bulk evolution.  Fig.~\ref{fig_YL} (left) shows results from this computation, demonstrating that the same-charge correlation data by STAR can be described quantitatively with CME and TMC together, computed with modest  magnetic field lifetime ($\sim 1\rm fm/c$) and realistic initial axial charge density.  The authors have made predictions  for the same-charge azimuthal correlations for various identified hadron paris: see  Fig.~\ref{fig_YL} (middle) for two-flavor case and Fig.~\ref{fig_YL} (right) for three-flavor case. The specific patterns with hadron identities  could provide highly nontrivial test when future data become available.  Correlations in certain channels   (e.g. $\gamma_{K^{+}K^{+}}$ and $\gamma_{p^+ \Lambda}$) are very sensitive to potential   contributions from strangeness sector.

Both studies, together in a complementary way, have conveyed a coherent message that the existence of  CME contribution is not only
 consistent with but also appears necessary for explaining the present 
 data. Further important progress is anticipated to come along this
 line of the quantitative CME study.

\subsection{CME-Related Phenomena}

Studying CME-related anomalous transport phenomena provides an independent avenue for the search of CME. We briefly discuss two examples: Chiral Magnetic Wave (CMW) and Chiral Vortical Effect (CVE). 

The Chiral Magnetic Wave is a gapless collective excitation (just like the sound wave) arising from intervening vector and axial charge fluctuations whose evolutions get mutually entangled by CME and CSE. The CMW is found to  induce an electric quadrupole, leading to  a rather specific splitting between the positive and negative pions' elliptic flow~\cite{Burnier:2011bf}:   $v_2^{\pm} \simeq  v^{\rm base}_{2, \pm} \mp r A_{\rm ch}/2$. A linear dependence of flow splitting $\Delta v_2 = v_2^{-} -v_2^{+}$ on event charge asymmetry $A_{\rm ch}$ has been unambiguously observed~\cite{Adamczyk:2015eqo}, with the extracted slope parameter $r$ roughly in agreement with CMW computations. Similar flow splitting measurements were done for $v_2$ and $v_3$ in AuAu and UU collisions, for identified pions and kaons, and across a variety of beam energy. All these data are  consistent with CMW expectations. A few possible background contributions~\cite{Steinheimer:2012bn,Bzdak2013,Voloshin:2014gja,Hatta:2015hca} were proposed but so far no alternative explanations were found to be compelling.  A recent study~\cite{Hatta:2015hca} proposed that the viscous transport of certain initial charge conditions may lead to sizable contributions to the observed flow splitting. While a number of aspects of this study may be questionable, it poses an important problem that should definitely be understood with realistic and quantitative simulations: namely, apart from possible {\it anomalous} transport via CMW, what would be the outcome of simply {\it normal} viscous transport of charges in heavy ion collisions.   The possibility that the observed flow splitting may be the result of CMW plus other effects, is of course still open and will need to be answered by future modelings.  On the experimental side, the so-called charge-multiple  analysis $Q^c_n$ (suggested a while ago in  \cite{Flow_CME,Liao:2010nv}) could provide very rich information for ``mapping out the charge landscape'' of the QGP in heavy ion collisions.

Global fluid rotation, quantified by the vorticity $\vec{\bf \omega} =  \bigtriangledown \times \vec{\bf v}/2$ (where $\vec{\bf v}$ is the fluid flow field), bears close similarity to an external magnetic field $\vec{\bf  B}$, as is evident even in classical physics between the Lorentz force  and the Coriolis force. Such similarity carries forward into quantum physics, and indeed similar to the CME, the vorticity induces a similar effect known as the CVE: $\vec{\bf J} = (1/\pi^2) \mu_5 (\mu \vec{\bf \omega})$ where $(\mu \vec{\bf \omega})$ (with $\mu$ the vector chemical potential) plays the role of $(e\vec{\bf B})$ in CME.  In typical non-central collisions there is nonzero angular momentum thus nonzero global rotation of the QGP fireball. Therefore just as in the case of CME, the (quark-level) CVE current leads to a separation of quarks and their anti-quarks across the reaction plane, with more quarks transported to one pole of the QGP fireball and more anti-quarks to the other pole. The CVE current  can thus manifest itself through the baryonic charge separation (in analogy to the electric charge separation of the CME)  which can be similarly measured through baryon-number-dependent azimuthal correlations $\gamma_{BB/{\bar{B}\bar{B}}}$ and $\gamma_{B\bar{B}}$~\cite{Kharzeev:2010gr}. In the most general case with both CME and CVE possibly occurring at quark levels, they would mix in very specific patterns for azimuthal correlations of various identified hadron pairs (see a more detailed discussion in \cite{Kharzeev:2015znc}). Such measurements have been reported recently by STAR~\cite{STAR_Wen}. The results, shown in Fig.~\ref{fig_data} (right), appear to suggest rather nontrivial ordering according to the hadrons' baryon number and electric charge. It would be highly interesting to see the extracted ``true'' signals in these various hadron pair channels after separating flow-driven component by analyses similar to those discussed in Sec. 3.2 for CME.

\subsection{Emergent New Ideas}

There are also many other interesting new ideas about CME and related physics, as reflected at this conference, which we could only briefly mention here due to limited space. By studying the interplay between a heavy obstacle and the CME current, a {\it Chiral Drag Force} has been found~\cite{Rajagopal:2015roa,Stephanov:2015roa}. In the presence of fluid rotation, a new gapless collective mode called {\it Chiral Vortical Wave (CVW)} is proposed in \cite{Jiang:2015cva} with its possible manifestation through elliptic flow splitting of $\Lambda$ and $\bar{\Lambda}$. Given that the strong magnetic fields may last only during the very early stage of heavy ion collisions, it is important to investigate possible early generation of CME in the pre-thermal glasma stage: such a study was reported recently in \cite{Fukushima:2015tza}. In addition to the CME and related anomalous transport phenomena, the extremely strong magnetic field should be able to manifest itself in other interesting ways. Indeed several proposal were made recently, including e.g.: possibly measurable Faraday and Hall effects~\cite{Gursoy:2014aka}; possibly enhanced elliptic anisotropy of charmonium production due to influence of early magnetic fields~\cite{Guo:2015nsa};   interesting interplay between magnetic helicity and the CME current~\cite{Hirono:2015rla}; etc.  This is only a very incomplete list with an emphasis on works reported at this conference, and we again refer the readers to recent reviews (e.g. \cite{Kharzeev:2015znc}) for further discussions.

\section{Summary}

In summary, the physics of Chiral Magnetic Effect (CME) and related anomalous transport phenomena is remarkably rich and exciting, with deep connections to QCD chiral symmetry, axial anomaly, and gluonic topology.   The study of CME has  led to important theoretical progress on developing  the anomalous hydrodynamics and the chiral kinetic theory, providing essential tools for quantitative modelings of anomalous chiral effects. It is of fundamental importance to search for the CME in heavy ion collision experiments. Recent achievements in separating flow backgrounds from CME signals, in the quantitative modelings of CME, as well as in measurements for CMW and CVE, have together presented a compelling evidence for the predicted effects. Of course a lot of works remain to be done in order to establish and quantify these effects and  to draw a final conclusion, which we hope would be accomplished in the near future.

\section*{Acknowledgements} 
The author thanks many colleagues for fruitful collaborations/discussions/communications on the topic of this contribution, in particular A. Bzdak, K. Fukushima, Y. Hatta, H. Huang, X.-G. Huang, Y. Jiang, D. Kharzeev, V. Koch,  L. McLerran, S. Mukherjee, M. Stephanov, A. Tang, S. Voloshin, F. Wang, G. Wang, N. Xu,  Z. Xu, H. Yee, and Y. Yin.    This work is  supported by the National Science Foundation (Grant No. PHY-1352368). The author is also grateful to the RIKEN BNL Research Center for partial support.





\bibliographystyle{elsarticle-num}
\bibliography{<your-bib-database>}



\end{document}